\documentclass[letterpaper,notoc]{JHEP3}
\usepackage{amsmath, amsfonts}
\usepackage{yfonts}
\usepackage{array}
\usepackage{cite}
\usepackage{graphicx}

\newcommand{\la}{\langle}
\newcommand{\ra}{\rangle}
\newcommand{\Tr}{{\rm Tr}}
\newcommand{\be}{\begin{equation}}
\newcommand{\ee}{\end{equation}}
\newcommand{\bea}{\begin{eqnarray}}
\newcommand{\eea}{\end{eqnarray}}
\newcommand{\nl}{\nonumber \\}

\newcommand{\bp}{{\bf p}}
\newcommand{\bq}{{\bf q}}
\newcommand{\bx}{{\bf x}}

\newcommand{\bv}{{\bf v}}
\newcommand{\dv}{\frac{d\Omega_v}{4\pi}}

\newcommand{\OO}{{\cal O}}

\newcommand{\Ref}[1]{(\ref{#1})}

\newcommand{\Nc}{N_{\rm c}}

\newcommand{\lsim}{\stackrel{\!<}{{}_\sim}}

\title{Hard thermal loops in the real-time formalism}

\author{Simon Caron-Huot\\
    McGill University, 3600 rue University,
    Montr\'eal QC H3A 2T8, Canada \\ E-mail: \email{scaronhuot@physics.mcgill.ca}}

\date {October 18, 2007}

\abstract{
We present a systematic discussion of Braaten and Pisarski's hard thermal
loop (HTL) effective theory
within the framework of the real-time (Schwinger-Keldysh) formalism.
As is well known, the standard imaginary-time HTL amplitudes
for hot gauge theory express the polarization
of a medium made out of nonabelian charged point-particles;
we show that the complete real-time HTL theory
includes, in addition, a second set of amplitudes
which account for Gaussian fluctuations in the charge distributions,
but nothing else.
We give a concise set of graphical rules which generate
both set of functions,
and discuss its relation to classical plasma physics.
}


\begin{document}

\section{Introduction}

It is a theoretically interesting problem to compute the higher-order
perturbative corrections received by physical observables in the quark-gluon
plasma,
especially in view of the present experimental program at RHIC \cite{star}
and of the future LHC heavy ion program.
In particular, one would like to understand theoretically
what the regime of validity
of perturbation theory is, for various observables of interest,
at least within the setup of a locally thermalized plasma.

Theoretical studies of the thermodynamic pressure, which is a quantity
naturally accessible to Euclidean space techniques, have revealed a poor
convergence of the perturbative series
unless the strong
coupling constant assumes unrealistically small values, $\alpha_s \lsim 0.1$.
In \cite{braatennieto} this behavior was attributed
to self-interactions of soft, $gT$ scale, gauge fields
($g\equiv\sqrt{4\pi\alpha_s}$).
Little is presently known about the corrections received
by \emph{dynamical} quantities, in particular those which are
leading-order sensitive to the $gT$ scale,
such as photon production rates \cite{photon1,photon2,photon3,photon4},
jet energy loss
\cite{jet1,photon4}, heavy quark energy loss \cite{heavy1,heavy2}
and transport coefficients (such as shear viscosity) \cite{transport}.
These quantities all are leading-order sensitive to the $gT$
scale, in the sense that they would
be logarithmically infrared divergent were the screening effects
which arise at this scale not properly resummed,
and are ``dynamical'' in the sense that they describe
real-time physics, making their extraction difficult from Euclidean
space correlators (and thus from lattice data).

The plasma effects which arise at the $gT$ scale are usually resummed,
in a gauge invariant way,
by means of Braaten and Pisarski's hard thermal loop (HTL)
effective theory \cite{braatenbig}.
This effective theory incorporates, to leading order in $g$,
the effects from the scale $T$ on the scale $gT$, in terms
of (nonlocal) effective propagators and vertices.
These ingredients can be used to build a loop expansion,
which, so long as only the soft scale $gT$ enters a problem,
is an expansion into powers of $g$.
Thus, since the above-listed dynamical quantities are leading-order
sensitive to the $gT$ scale,
one expects them to receive potentially
large $\OO(g)$ corrections from soft loops in the HTL theory.

Until recently, there had been no calculation of the $\OO(g)$
corrections received by \emph{any} of these quantities.
We believe there is a certain technical advantage in performing
such calculations directly in Minkowski space,
using real-time (Schwinger-Keldysh) techniques; however to the best
of our knowledge a systematic discussion of the HTL theory within
this formalism is presently lacking from the literature.
In this paper we provide just such a discussion.
We have recently applied the theory we present here
to a next-to-leading order calculation
of nonrelativistic heavy quark diffusion \cite{ourpaper}.

This paper is organized as follows. In section \ref{sec:realtime}
we briefly recall the rules of the real-time formalism, in particular
within the so-called Keldysh (``$r/a$'') basis. 
In section \ref{sec:power} we show, by a power-counting argument,
that the HTL effective theory
takes on an especially simple form in this basis: the only HTL
amplitudes (with external gauge bosons)
carry at most two $a$ Keldysh indices, though arbitrarily
many $r$ indices.
In section \ref{sec:compute}
we compute these amplitudes, which have a simple and physically
transparent form, and give their generating functional.

A convenient set of effective graphical rules which generate
the real-time HTL theory is given in section \ref{sec:rules};
these rules are essentially a graphical realization of the
nonabelian Vlasov equations (including, as well, Gaussian fluctuations
in the particle distribution functions.)
We discuss the relationships between our results and previous work,
and with the classical plasma physics of a gas of point-like
nonabelian charges, in section \ref{sec:discuss}.
Since our method of analysis appears to shed little light on the structure
of real-time perturbation theory when soft fermions are involved,
we leave the analysis of fermionic HTLs to future work.

\section{The real-time formalism}
\label{sec:realtime}

The real-time formalism allows the
description of the dynamical evolution of expectation values
within some initial state or density matrix
(as opposed to ``in-out'' transition amplitudes).
The formalism is characterized by a doubling of the degrees of
freedom:
in addition to the usual ``$\phi_1$'' fields which implement forward
time evolution, one should path-integrate over a second set of
fields, ``$\phi_2$'', which implement time evolution backward in time
to some initial time. We work in the so-called
Keldysh $r/a$ basis, obtained via the change of
basis $\phi_r=\frac12(\phi_1+\phi_2)$ and $\phi_a=\phi_1-\phi_2$,
and let the initial time at which the system's density matrix is defined
go to $-\infty$.
For a review we refer the reader to \cite{Chou}%
\footnote{Our $r/a$ fields correspond to the $1/2$ fields of the
  ``physical representation'' used by these authors. }.

Here we merely recall the rules of perturbation theory in this context.
In thermal equilibrium the propagator is a
$2\times 2$ matrix, which takes the form
\be G \equiv \left(\begin{array}{cc} G_{rr} & G_{ra} \\ G_{ar} & G_{aa}
\end{array}\right) =
\left(\begin{array}{cc} \displaystyle
(G_R-G_A)\left(\frac12 \pm n(p^0)\right) & G_R \\ G_A & 0 \end{array}
\right) \label{racorrel},
\ee
where $n(p^0)\equiv1/(e^{p^0/T}\mp 1)$
denote the standard Bose-Einstein and Fermi-Dirac
distributions, for bosons and fermions
respectively. For a free scalar field,
the retarded propagator $G_R$ would be given as%
\footnote{Our metric is $({-}{+}{+}{+})$, and following
  finite temperature conventions, we capitalize four-vectors but write
their components as lowercase.
}
$G_R(P)=-i/(P^2+m^2-i\epsilon p^0)$ in Fourier space.
The general form \Ref{racorrel} holds nonperturbatively:
the propagator is completely determined by $G_R$.
The chief reason for using the $r/a$ basis in this work
is that Bose-Einstein distributions, which play
a key role when treating soft physics, appear in only one
matrix element of the propagator,
and are therefore most conveniently managed.

To perform perturbative calculations, one must
sum over the $r/a$ assignments for all internal legs in Feynman diagrams,
subject to the restriction that the vertices
carry an odd number
of $a$ indices. The vertices having one $a$ index coincide with
the standard zero-temperature ones, and those having three $a$ indices are smaller
by a factor $1/4$.
External $r$ and $a$ fields in correlation functions carry distinct physical meaning:
since the difference field $\phi_a$ is analogous to
an interaction term which would be added to the Hamiltonian,
general correlators of $a$ and $r$ fields may be understood
in terms of the (retarded) nonlinear response induced by the $a$ fields
on some correlator of $r$ fields \cite{Chou}.   
Correlators in which the $a$ field has the largest time argument
vanish, $\phi_a(t)\to 0$ as $t\to \infty$.

We graphically represent Feynman diagrams in the $r/a$ formalism by
drawing incoming arrows on $r$ fields which enter interaction vertices,
and outgoing arrows on $a$ fields.
With this notation, retarded and advanced propagators
bear a single arrow, which points in the direction of
the flow of time. An $rr$ propagator carries two outgoing arrows,
which we separate by a ``cut'': we think of the cut as a place
where the time flow can start. Sometimes we will omit to draw the
arrows on these propagators, which should create no confusion.
The (tree-level) interaction
vertices all have an odd number of outgoing arrows.
Our graphical notation is illustrated in Fig.\ \ref{fig:examplera}.
Incidentally, this particular diagram yields zero,
because it contains a closed loop of retarded propagators.

\begin{FIGURE}[ht] {
\centerline{\includegraphics[width=10cm,height=2.5cm]{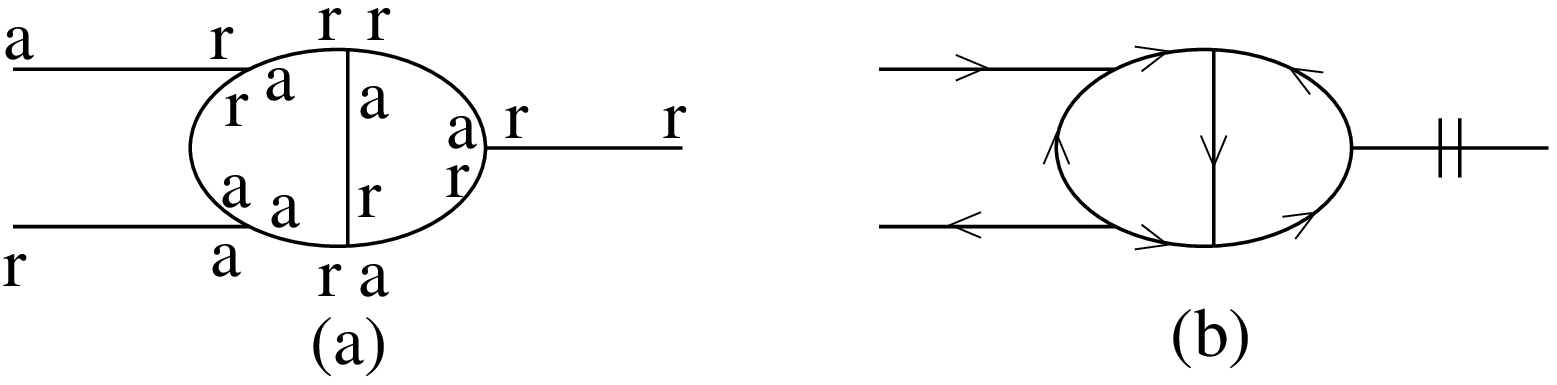} }
\caption{Example of a Feynman diagram in the $r/a$ formalism, with
(a) explicit marking of $r/a$ indices and (b) our graphical
  notation. The propagators which carry arrows are retarded, and the cut
propagator is an $rr$ propagator.}
\label{fig:examplera}
} \end{FIGURE}

\section{Power counting}
\label{sec:power}

We consider (amputated) vertex functions in which all of the
external legs are soft gauge bosons; by soft we mean $P\sim gT$
and by hard we mean $P\sim T$, for all components of $P$.
We recall that vertex functions having only one Keldysh $a$ index, often
called ``fully retarded functions'',  correspond
to a direct analytic continuation of the Euclidean vertex functions
\cite{Evans}, and that these HTL amplitudes are of
parametric size $g^2T^{4-n}$,
where $n$ is the number of external legs \cite{braatenbig,frenkeltaylor}.
Amputated vertex functions with no external $a$ index vanish, because
correlation functions involving only $a$ fields vanish.
There are no HTL amplitudes involving external ghost fields \cite{braatenbig},
at least within the classes of covariant and Coulomb gauges,
so we will not consider diagrams with external ghost lines.
We now show that loop amplitudes with $n$ external legs, $n_a$ of which
bearing Keldysh $a$ indices,
can only compete with the above-mentioned HTL amplitudes if they
are of parametric size $g^{3-n_a}T^{4-n}$. Then we
show that no hard loop can be parametrically larger than $gT^{4-n}$,
implying that vertex functions with $n_a\geq 3$ are not part of the HTL
theory. We also show that the soft contribution to one-loop amplitudes
behaves like $g^{4-n_a}T^{4-n}$ and produces subleading effects relative
to the HTLs, although vertex functions with $n_a\geq 3$ are
soft-dominated, not hard-dominated. The key ingredients entering our
power-counting are summarized in Table \ref{tab:ingredients}.

Vertex functions with more external Keldysh $a$
indices, when they appear within
Feynman diagrams, tend to be suppressed due to
the (absence of) Bose-Einstein factors on the propagators connecting to them.
To see this, we note that
when a vertex function carries an $a$ index, the Keldysh index
on the remote side of the propagator connecting to it must necessarily
be an $r$ index, because of the absence of an $aa$ propagator.
There thus automatically exists a corresponding diagram in
which the $a$ index on the vertex function is replaced with an $r$,
and the $ar$ propagator replaced with an $rr$ propagator;
since for soft external momenta the latter propagator
is larger by one Bose-Einstein factor $T/p^0\sim
1/g$, we see that the original vertex function will only compete
with the latter if it is parametrically larger by one factor $1/g$.
By induction this proves our first claim:
amputated vertex functions with $n$ external
legs, $n_a$ of which carrying $a$ indices, will only compete with
the HTL amplitudes with one external $a$ index
if they are of parametric size $g^{3-n_a}T^{4-n}$.

\begin{TABLE} {
\begin{tabular}{|l|c|} \hline
Ingredient & Parametric strength \\ \hline
Soft (bosonic) retarded propagator & $1/g^2T^2$ \\
Soft (bosonic) $rr$ propagator & $1/g^3T^2$ \\
Hard, near light-like propagator & $1/gT^2$ \\
Hard three-point vertex & $gT$ \\
Soft three-point vertex & $g^2T$ \\
$d^4Q$ for hard, near light-like $Q$ & $gT^4$ \\
$d^4Q$ for soft $Q$ & $(gT)^4$ \\ \hline
\end{tabular} 
\caption{Ingredients which enter our power-counting.}
\label{tab:ingredients} }
\end{TABLE}

We now show that it is impossible for a hard loop with $n$ soft external
(gauge boson) legs to be parametrically larger than $gT^{4-n}$.
Indeed, let us consider a bosonic loop diagram
(the conclusion being unchanged for fermionic loops),
having $n$ three-point vertices
and no four-point vertices (it can be shown that the latter get
suppressed in general.)
The dominant contribution from the region of hard loop momentum $Q\sim T$
arises when $Q$ is within $gT$ of the light cone, in which case
each propagator contributes a large $\sim 1/gT^2$ factor.
This is because, for generic soft external momenta,
no two hard propagators
can simultaneously become closer to the light cone than $gT$,
and parametrically nothing special happens when only one propagator
becomes arbitrarily close the the light-cone
(because of the corresponding measure suppression); this region
maximizes the number of simultaneously large propagators.
The estimate $1/gT^2$ for hard, near light-like, propagators
is independent on the hard propagator being retarded or cut%
\footnote{Enforcing the mass-shell condition
on some hard propagators, when they are $G_{rr}$
propagators, can force linear combinations of the external momenta
to be spacelike, but this does not
represent a \emph{parametric} suppression of the phase space of
the soft external momenta.
What we are saying is that, for $Q=R+P$
with $R^2=0$ and $P$ soft, $Q^2\approx 2R\cdot P$, it is fair to treat
$1/(2R\cdot P\pm i\epsilon r^0)$
and $\delta(2R\cdot P)$ as parametrically equivalent, $\sim 1/gT^2$. }.
The restriction of the integration measure $d^4Q$
to the hard, near light-like region produces a factor of $gT^4$,
and counting the $n$
three-point interaction vertices each as $gT$,
we find that a hard loop with $n$ external
legs can behave at most like $gT^{4-n}$ parametrically.
Actually a cancellation occurs when $n_a=1$ so these functions
behave like $g^2T^{4-n}$, but we will see in the next section
that no such cancellation occurs when $n_a=2$.
Together with the last paragraph, this shows that \emph{hard}
loops with $n_a\geq 3$, when appearing inside
full diagrams, have subleading effects relative to those with $n_a=1,2$.

What about soft loops?
Using the rules of the $r/a$ formalism, it can be shown
that loop diagrams with $n_a$ external $a$ indices
can contain up to $n_a$ internal cut ($rr$) propagators.
For bosonic loops, the presence of $n_a$ Bose-Einstein functions
suggests that diagrams with $n_a$ large
should be soft-dominated, not hard-dominated.
This indeed happens: in the soft region,
a (bosonic) retarded propagator should
be estimated as parametrically $1/g^2T^2$, a cut propagator
estimated as $1/g^3T^2$, a three-point interaction vertex treated
as $g^2T$, and the integration measure treated as $(gT)^4$.
Adding up, we find that the soft loop contribution to a
bosonic loop diagram is parametrically $g^{4-n_a}T^{4-m}$.
Since soft loops should actually be evaluated using effective
HTL-resummed propagators and vertices, which are quite complicated
expressions, we do not expect parametric cancellations to occur.

For $n_a=1,2$ the soft contribution is down by exactly one power
of $g$ relative to the hard contribution: these diagrams
are truly hard-dominated and deserve to be called ``hard thermal loops.''
Hard loops with
arbitrary external $r/a$ indices were considered in \cite{fueki01},
in which it was found that due to a cancellation hard loops
with $n_a=3$ were of parametric order $g^2T^{4-m}$
(however these authors did not recognize that hard loops with $n_a\geq 3$
only had subleading importance in actual calculations).
Thus we see that hard loops with $n_a\geq 3$ not only
have subleading effects, they are also incorrect: the
corresponding diagrams
are \emph{soft}-dominated, not hard-dominated.
However this poses no problem to power-counting: when present inside
diagrams, soft loops having arbitrary numbers of external $a$
indices all contribute at the same order as the soft
contribution to the $n_a=1$ loops, e.g. are down by $\OO(g)$
relative to the HTL contribution. Essentially what happens is that by
modifying the external $r/a$ indices on a soft loop,
one is merely transferring
the Bose-Einstein factors back and forth between the propagators
outside and inside the loop.

The HTL amplitudes account for the dominant effects from hard particles
on soft gauge fields, and together with the tree interactions vertices
(which are of the same order),
they can be used to set up an effective theory which contains only the $gT$
scale.
We believe that the loop expansion within this theory, in the real-time formalism,
proceeds in a way entirely similar to that in the imaginary-time formalism \cite{braatenbig}: 
as long as only the scale $gT$ contributes,
each additional loop is suppressed by one power of $g$ %
\footnote{Of course, in practice such an expansion is not expected
to hold up to arbitrarily high order, since other scales should
eventually enter (associated e.g. with some mean free path,
or with nonperturbative, infrared physics).
However, the appearance of a new physical scale would
be signaled by a divergence in the HTL effective theory, because
it contains only one scale.}. In general, in a theory with only one physical
scale, one would expect the loop expansion to be an expansion into
powers of $g^2$; in the imaginary-time formalism
one gets an expansion into powers of $g$, because
each additional soft loop can introduce one (and only one) additional
Bose-Einstein factor.
Within the real-time formalism, this well-known
statement must be modified to the 
claim that each additional loop introduces one (and only one)
additional Bose-Einstein factor \emph{or} HTL vertex functions with two external $a$ indices.
The latter vertices may be regarded as secretly containing one
Bose-Einstein factor, since they are larger by a factor $1/g$.
We will not embark here into a general proof of this claim,
which would be easy to give using the effective Feynman rules
we present in section \ref{sec:rules};
we simply remark that it would be quite surprising for a systematic
loop expansion to hold within one formalism, and not within another.

\section{Calculation of the hard thermal loops}
\label{sec:compute}

\subsection{HTLs with one $a$ index}

The gluonic hard thermal loops having only one external Keldysh $a$ index,
often called ``fully retarded functions'',
can be obtained via a direct analytic continuation of the
well-known Euclidean ones \cite{braatenbig} \cite{frenkeltaylor},
and thus do not need to be independently recomputed within
the real-time formalism.
Nevertheless, we think it is instructive to briefly
describe the Feynman diagrams which contribute to them,
and how, following \cite{blaizotkin,jackiwnair}
they can be evaluated using
a simple kinetic theory of point-like particles.

\begin{FIGURE}[ht] {
\centerline{\includegraphics[width=4.7cm,height=2.6cm]{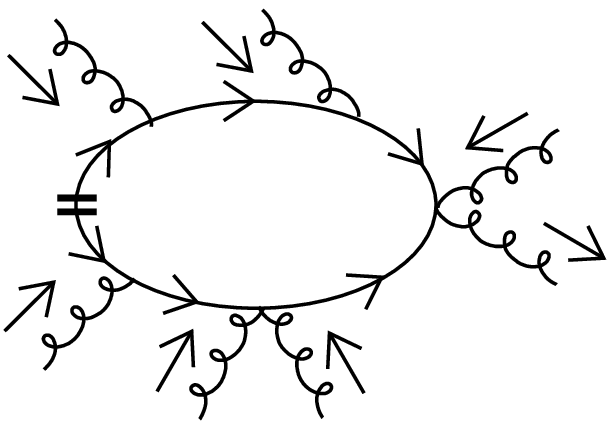}}
\caption{
The most general one-loop diagram with one external $a$ index
(the outgoing arrow),
but many $r$ indices: all such diagrams only have one cut propagator.
The arrow flow corresponds to time flow;
we do not specify the nature of the particle running in the loop.}
\label{fig:htl1}
} \end{FIGURE}

The most general one-loop Feynman diagram
with one external $a$ index, but arbitrarily many $r$ indices,
is illustrated in Fig.\ \ref{fig:htl1}. All of these
diagrams contain only one cut ($rr$) propagator, which occurs
at the smallest time in the diagram.
In terms of our graphical notation, this structure can be
tracked down to the fact that inserting an external
$r$ field (incoming arrow) onto a diagram (involving
only bare interaction vertices),
be it onto an existing vertex or onto a propagator,
never modifies its arrow flow.
Physically, this structure means
that these functions represent the nonlinear response
of the expectation value of the gauge current (at the $a$ leg)
due to a background field (the $r$ legs):
the Feynman diagrams mimic
the obvious quantum-mechanical procedure
of starting with an initial one-particle density matrix
(here, the ``cut''),
evolving it under a background field, and
taking the expectation value of the current at
late times.

In \cite{blaizotkin} the problem of calculating the induced
current in a soft background gauge
field (mean field) was considered, and reduced to kinetic theory:
\bea
J^{\mu\,a}_{\rm ind}(X) &=& \sum_{\rm DOF} g
\int \frac{d^3p}{(2\pi)^3} \,v^\mu n^a(X,\bp), \label{systemarr1} \\
v\cdot D\, n^a(X,\bp) &=& g\Tr_r\left[t^a_r t^b_r\right]
dv^i E^{i\,b}(X) \frac{-dn(p)}{dp}
\label{systemarr2},
\eea
where $v^\mu=(1,\bp/p)$ represents the four-velocity of a hard particle
and $n$ denote the standard Bose-Einstein or Fermi-Dirac distributions.
The second equation, to be solved with retarded boundary conditions,
determines the color-adjoint disturbance $n^a$.
The concept of point-like particles originates from
the separation of scales $gT\ll T$, between the momenta of the
external gauge field and that of the typical particles which
contribute to the induced current: the hard particles
feel the external field as if they were point particles.
The degree of freedom count in \Ref{systemarr1} is: two bosonic degrees
of freedom for the gauge/ghost system, four fermionic degrees of freedom
for Dirac fermions, and two bosonic degrees of freedom for complex
scalar fields; the form of the resulting equations is the same
for all of these particles, and is gauge-fixing independent \cite{braatenbig}.
A collision term is not included in \Ref{systemarr2} because collisions are
only relevant over $1/g^2T$ time scales;
for a more ample discussion we refer to the review \cite{blaizot01}.

Solving for the induced current \Ref{systemarr1}-\Ref{systemarr2} with retarded boundary
conditions yields the term
in the generating functional (effective action) of
real-time amplitudes which is linear in the Keldysh $A_a$ field:
\bea \hspace{-0.6cm} \Gamma^{(1)} &=& 
m_D^2 \int\dv \int d^4X \,v\cdot A_a
\frac{1}{v\cdot D[A_r]} \bv\cdot{\bf E}[A_r] \nl &\equiv&
m_D^2\int \dv\int d^4X \int_0^\infty d\tau\,
v_\mu A_a^{\mu\,a}(X) \,U^{ab}(X,X-v\tau)[A_r]\, v^i E^{i\,b}[A_r](X-v\tau)\,,
\label{gena}
\eea
where $U^{ab}$ stands for an adjoint Wilson line
along the hard particle trajectories. Here we have explicitly performed
the radial integration in \Ref{systemarr1}, leaving
only the integration over the angle $\bv$ of the hard particles;
in a generic Yang-Mill
theory with $N_F$ Dirac fermions and $N_S$ complex scalars,
the degree of freedom counts add up to:
\be m_D^2=\frac{g^2T^2}{3}\left[ C_A + N_F T_F +N_S T_S \right]\,, \label{defmd2}
\ee
where $C_A=\Nc$ 
and $T_F=T_S=\frac12$ in SU($\Nc$) gauge theory with  matter
in the fundamental representation.

We find \Ref{gena} rather physically transparent
compared to its Euclidean counterpart
\cite{braatengen} \cite{taylorwonggen},
however compact the latter might be.
The equivalence between the analytic continuation
of the vertex functions derived from \Ref{gena} and
the standard Euclidean ones can be verified from
the explicit expressions for the induced
current $\delta \Gamma_{\rm Euclidean}/\delta A$ given in \cite{taylorwonggen}.
It is a rather nontrivial, although necessary, property
of these functions that they become symmetrical
in all of their arguments (including the $a$ leg)
when the boundary conditions are made symmetrical, by analytically
continuing the momenta to imaginary frequencies.
However, it is this very asymmetry
between the $a$ leg (``induced current'') and the $r$ legs
(``external fields'') of the fully retarded vertex functions,
at physical values of the momenta,
which makes the generating functional \Ref{gena} so simple.
We describe the Fourier space amplitudes
derived from \Ref{gena} in more detail, in section \ref{sec:rules}.

\subsection{HTLs with two $a$ indices}
\label{sec:computeaa}

We now compute the hard thermal loops with two external $a$
indices, beginning with the $aa$ self-energy.
The relevant Feynman diagrams
are shown in Fig.\ \ref{fig:htl2} (a)-(c); the propagators in
these diagrams are most conveniently added together under the
integration sign:
\be
G_{rr}(Q) G_{rr}(R) + \frac14 G_R(Q) G_R(R) + \frac14 G_A(Q) G_A(R)
\simeq \frac12 \left[ G^>(Q)G^>(R) + G^<(Q)G^<(R) \right] \label{toaacut}\,,
\ee
the equality holding up to analytic terms which integrate to zero
such as $G_R(Q)G_A(R)$ (``closed loops of retarded propagators''),
which we have subtracted in passing
to the right-hand side. Here the propagators $G^{>,<}$ denote
the Wightman (unordered) two-point functions,
and we have employed the identities $G_R-G_A=G^>-G^<$,
$G_{rr}=\frac12(G^>+G^<)$.
Equation \Ref{toaacut} states that the $aa$ self-energy
is the average of the two Wightman self-energies, a cutting
pattern which we graphically represent with two parallel lines,
as in Fig.\ \ref{fig:htl2} (d),
in analogy with our notation for the $G_{rr}$ propagator.
Evaluating the loop for a complex scalar field, for definiteness, yields:
\bea
i\Gamma_{aa}{}^{ab}_{\mu\nu}(P) &=& -g^2 \Tr_r \left[t^a_r t^b_r\right]
\int \frac{d^4Q}{(2\pi)^4} \,(2Q+P)_\mu
(2Q+P)_\nu
\frac{G^>(Q)G^>(R) + G^<(Q)G^<(R)}{2}
\nl &\approx& -g^2\sum_{\rm DOF} \Tr_r \left[t^a_r t^b_r\right]
\int \frac{d^3q}{(2\pi)^3} 2\pi\delta(v\cdot P)\,v_\mu v_\nu \,n_B(q)(1+n_B(q))\,,\label{systemaa1}
\eea
where on the second line we have used $P\ll Q$ since the integral
is saturated for $Q\sim T$.
We see that the calculation of this self-energy is relatively simple,
compared to the corresponding retarded HTL self-energy:
since no $\OO(g)$ cancellation
occurs, all $P\ll Q$ approximations can be applied directly.
Particles of different spins yield similar contributions (up to $\OO(g)$
corrections), and one gets the same degree of freedom count
as in \Ref{systemarr1}.
The self-energy \Ref{systemaa1} could also be obtained by means of the KMS
(fluctuation-dissipation)
relation, which relates the $aa$ HTL self-energy to $T/p^0$ times the discontinuity
of the retarded HTL self-energy.

\begin{FIGURE}[ht] {
\includegraphics[width=13cm,height=2.3cm]{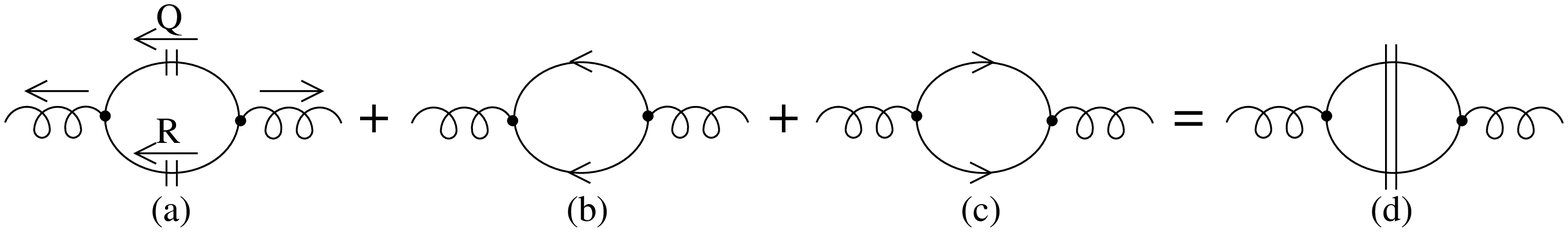}
\caption{(a)-(c): Diagrams contributing to the $aa$ gluon self-energy,
with momentum assignments displayed on the first diagram.
(d) The sum of the three preceding diagrams.
The double
cut represents the average of the Wightman cuts in the two direction.}
\label{fig:htl2}
} \end{FIGURE}

Equation (\ref{systemaa1}) has a rather straightforward physical interpretation:
the double cut in diagram (d) describes
thermodynamical equal-time fluctuations,
\be \la n^a(\bx,\bq) n^b(\bx',\bq') \ra =
\Tr_r\left[t^a_r t^b_r\right]
\delta^3(\bx-\bx')(2\pi)^3 \delta^3(\bq-\bq')\,n_B(q)(1+n_B(q))\,,
\label{fluctaa}
\ee
affecting the distribution functions which appear in \Ref{systemarr1}.
Expression \Ref{systemaa1} is precisely $(i)^2$ times
the Fourier transform of the unequal time
correlator derived from \Ref{fluctaa}, by means of ballistic propagation
$v\cdot \partial n^a(X,\bq)=0$. 
Considerations of gauge invariance
immediately suggest that to obtain the fluctuation functions with external
$r$ gauge fields, one should merely substitute this
free ballistic propagation with a gauge-covariant one, $v\cdot
D[A_r]n^a(X,\bq)=0$.

\begin{FIGURE}[ht] {
\centerline{\includegraphics[width=4.5cm,height=2.4cm]{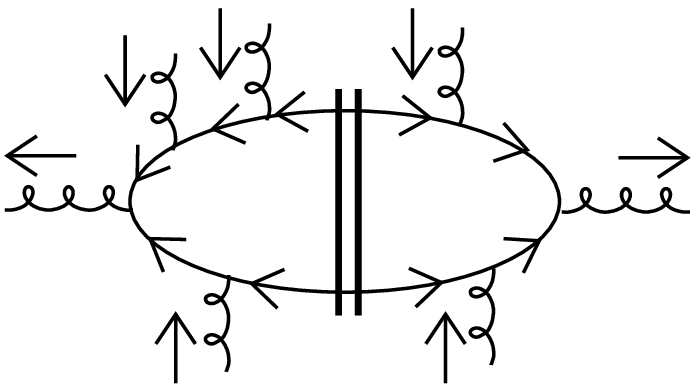}}
\caption{The most general form of diagram contributing to gluon HTLs with
two $a$ indices but many $r$ indices.
Diagrams with four point vertices are subleading. The double-cut has the same
meaning as in Fig.\ \ref{fig:htl2}. }
\label{fig:htl3}
} \end{FIGURE}

Not surprisingly, this expectation is borne out by explicit calculations.
This is especially obvious to see if one begins
with a reorganization of the $r/a$ structure of the relevant diagrams,
in a way analogous to \Ref{toaacut} above.
Indeed, although a direct application of the
rules of the Schwinger-Keldysh formalism would yield diagrams
which correspond to inserting the external $r$ fields onto
the diagrams (a)-(c) of Fig.\ \ref{fig:htl2}, it can be proved
that their sum is equivalent to what is obtained by inserting
the $r$ fields directly onto the simpler diagram (d):
all of the diagrams thus obtained contain two cut (Wightman)
propagators, which appear at the smallest time,
as illustrated in Fig.\ \ref{fig:htl3}.
Making all small $P$ approximations to the propagators and vertices
in these diagrams amounts to taking the propagation of the hard
particles to be eikonal, e.g. given by Wilson lines along
their classical trajectories, justifying
the procedure mentioned in the preceding paragraph of replacing
$v\cdot \partial$ with $v\cdot D[A_r]$.

Thus, upon performing the radial integration in \Ref{systemaa1},
we obtain the generating functional for vertex functions with
two $a$ indices (with $m_D^2$ as in \Ref{defmd2}):
\be \Gamma^{(2)}
=\frac{m_D^2 i T}{2} \int \dv \int d^4X \int_{-\infty}^\infty d\tau\,
v_\mu A_a^{\mu\,a}(X) \,U^{ab}(X,X-v\tau)[A_r]\, v_\nu A_a^{\nu\,b}(X-v\tau)
\label{genaa}
\ee

Although it should be rather obvious, from their manifest gauge-covariance,
that the vertex functions obtained from \Ref{gena} and \Ref{genaa}
obey a large class of Ward identities, we remark that the full
HTL effective action $\Gamma^{(1)} + \Gamma^{(2)}$
is not strictly invariant under the whole set of Schwinger-Keldysh
gauge transformations%
\footnote{More precisely it is invariant
under those infinitesimal gauge transformations whose parameter
is a Keldysh $r$ field, but not under
those for which it is an $a$ field.
}.
Such a strict invariance would require the inclusion of
terms which are $\OO(A_a^3)$ and higher order in the Keldysh $A_a$ fields,
which, however, we prefer not to include since
we argued in section \ref{sec:power} that such terms
are not part of the HTL effective theory.

\section{Feynman rules for kinetic theory}
\label{sec:rules}

We now present a concise set of Feynman rules 
which generates the complete (gluonic) HTL effective theory
in the real-time formalism.
To obtain these rules we first rewrite the
induced current $\delta\Gamma/\delta A_a$ from \Ref{gena} into the form:
\be J^{\mu\,a}_{\rm ind}(X)
= m_D^2\left[ -A^0(X)\delta^\mu_0 + \int \dv \frac{1}{v\cdot D} \partial^0
v\cdot A^a\right] \label{j0ind2},
\ee
where we have decomposed
the electric field as $E^i=\partial^0 v\cdot A - v\cdot D A^0$
and used $\int \dv v^\mu=\delta^\mu_0$.
Expressions of the form $(1/v\cdot D) S$ should be understood
as the solution of $v\cdot D=S$ with retarded boundary conditions,
e.g. the adjoint Wilson line in \Ref{gena}.

\begin{FIGURE} {
\renewcommand{\u}{\unitlength}
\begin{picture}(365,83)(-30,0)
\put(0,50){\includegraphics[width=60\u,height=19\u]{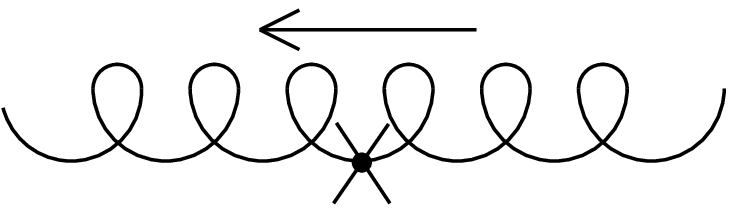}}
\put(0,25){\includegraphics[width=60\u,height=16\u]{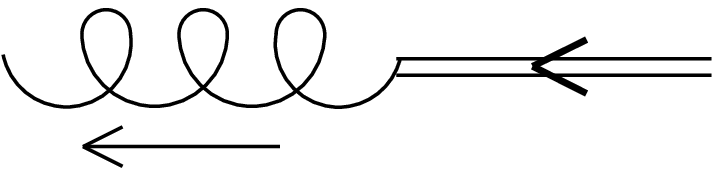}}
\put(0,3){\includegraphics[width=60\u,height=16\u]{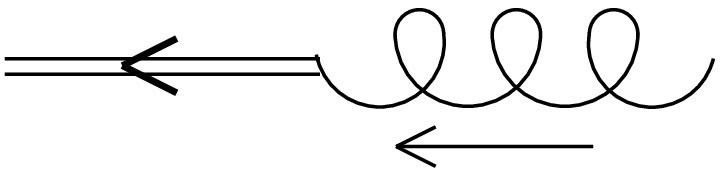}}
\put(-30,55){(a)}
\put(-30,31){(b)}
\put(-30,8){(c)}
\put(-8,60){$\mu$}
\put(62,60){$\nu$}
\put(-8,35){$\mu$}
\put(62,13){$\mu$}
\put(10,1){$P$}
\put(75,56){$= im_D^2 \delta^\mu_0\delta^\nu_0$}
\put(75,34){$= iT\,v^\mu$}
\put(75,12){$= ip^0\,v^\mu$}

\put(220,55){\includegraphics[width=40\u,height=22\u]{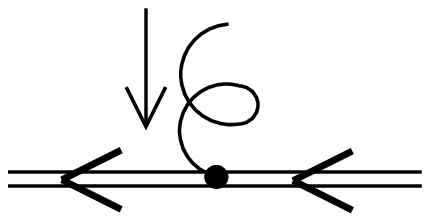}}
\put(238,78){$\mu,b$}
\put(213,58){$a$}
\put(261,58){$c$}
\put(272,56){$=-v^\mu f^{abc}$}
\put(220,29){\includegraphics[width=40\u,height=8\u] {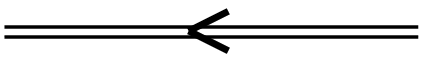}}
\put(235,38){$P$}
\put(272,32){$\displaystyle = \frac{-i}{v\cdot P^-}$}
\put(220,06){\includegraphics[width=40\u,height=10\u]{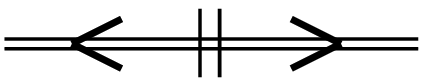}}
\put(235,16){$P$}
\put(272,08){$= 2\pi\delta(v\cdot P)$}
\put(190,55){(d)}
\put(190,31){(e)}
\put(190,08){(f)}
\end{picture} 
\caption{Effective Feynman rules for the HTL theory in the $r/a$ formalism.
The arrows follow the graphical notation for $r/a$ diagrams introduced in section
\ref{sec:realtime}.
All two-point functions are proportional
to the identity in color space, $\delta^{ab}$, not explicitly shown.
A factor $(m_D^2/T)\int\dv$ must be given to every disjoint double line
appearing in a diagram.}
\label{fig:htlrules}
} \end{FIGURE}

The graphical rules given in Fig.\ \ref{fig:htlrules} reproduce this
induced current, which gives the amplitudes with only one external $a$ index
(outgoing arrow).
Specifically, the $A^0\delta^\mu_0$ term in \Ref{j0ind2}, which
gives rise to a ``contact term'' in the retarded
self-energy, is mapped to the component (a) of Fig.\ \ref{fig:htlrules}.
The second, non-local term in \Ref{j0ind2} is mapped to a class of
diagrams, in which an incoming gluon first generates
a disturbance in the distribution function associated with the
four-velocity $v^\mu$ via vertex (c),
which is then evolved using the eikonalized retarded propagator
(e) and interaction vertex (d). The current associated with this disturbance
sources a gauge field via vertex (b).
Finally, one must perform the integration over the disturbed particle's
momentum by adjoining a factor $m_D^2 \int \dv$ to all double lines
in a diagram. The cut eikonal propagator (f), which plays no role in the
calculation of the HTL amplitudes with only one external $a$ index,
enters the calculation of amplitudes with two external
$a$ indices (in which it appears exactly once). By ($v\cdot P^-$)
we mean
\be \frac{1}{v\cdot P^-} \equiv \frac{1}{v\cdot P-i\epsilon}\,.
\ee

The effective propagators and interaction vertices 
of Fig.\ \ref{fig:htlrules} are to be used in building
Feynman diagrams according to the standard rules of the $r/a$
formalism; the objects (a)-(c) correspond to
$ar$ interaction vertices, the propagator (e) is a retarded ($ra$)
propagator and (f) represents an $rr$ propagator.
The interaction vertex (d) carries $arr$ indices and is the only
such interaction vertex in this theory; there is no three-point
vertex involving an $a$ gluon.
In practice the self-energies on soft gluon lines must be resummed;
when this resummation is performed the insertion (a) should be ignored,
as well as all diagrams in which only two gluons connect to a double line.
We have not explicitly shown the tree interaction vertices between
soft gluons, although these are of the same order as the
HTL ones and must be included in the effective theory.

The double lines in our graphical
rules may be thought of as two-particle states
(alternatively, one-particle density matrices):
in general, by ``opening up'' these double lines the
one-loop diagrams considered in section \ref{sec:compute}
are recovered, or more precisely, specific sums of these diagrams.
As we comment on below, these graphical rules represent
purely classical plasma physics.

\subsection{An example}

\begin{FIGURE} {
\renewcommand{\u}{\unitlength}
\begin{picture}(300,60)(0,0)
\put(0,10){\includegraphics[width=300\u,height=40\u]{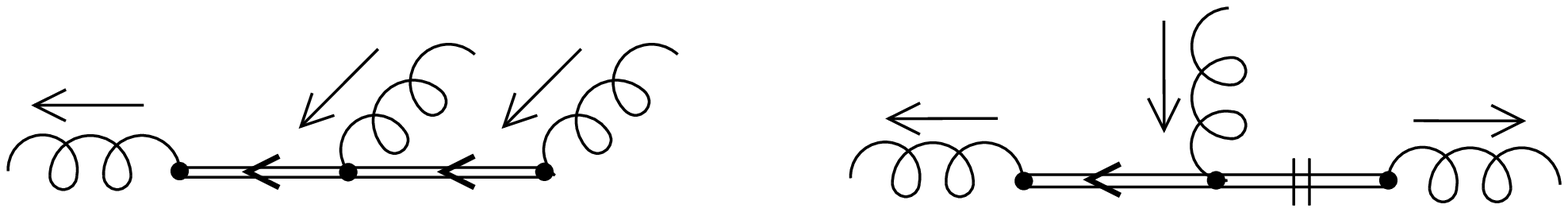}}
\put(58,0){(a)}
\put(225,0){(b)}
\put(0,35){$P,\mu,a$} 
\put(60,45){$Q,\nu,b$} 
\put(105,45){$R,\sigma,c$} 
\put(165,35){$P_1,\mu,a$}
\put(262,35){$P_2,\nu,b$}
\put(215,53){$Q,\sigma,c$}
\end{picture}
\caption{Effective Feynman diagrams contributing to HTLs
with three external legs, up to permutations. (a) HTLs with
one $a$ index. HTL. (b) HTLs with two $a$ indices.}
\label{fig:htlexample}
} \end{FIGURE}

As an example of the application
of these rules, we evaluate the three-point HTLs.
The relevant diagrams are shown in Fig.\ \ref{fig:htlexample},
and upon adding their permutations, the application
of the rules produces:
\bea \Gamma_{arr}{}^{abc}_{\mu\nu\sigma}(P;Q,R) &=& i m_D^2 f^{abc}\int \dv
\frac{v_\mu v_\nu v_\sigma}{v\cdot P^-} \left[ \frac{r^0}{v\cdot R^-} -
\frac{q^0}{v\cdot Q^-}\right], \label{ex:gammaarr} \\
\Gamma_{aar}{}^{abc}_{\mu\nu\sigma}(P_1,P_2;Q) &=& m_D^2 T f^{abc}
\int \dv \frac{v_\mu v_\nu v_\sigma}{v\cdot Q^-} \left[ 2\pi\delta(v\cdot
P_2) - 2\pi\delta(v\cdot P_1)\right]. \label{ex:gammaaar}
\eea
Our notation for $\Gamma$, which corresponds to $(-i)$ times the
amputated Feynman
diagrams themselves, is that the momenta written before the semicolon
are outgoing and the others are incoming. In particular, momentum
conservation implies $P=Q+R$ and $P_1+P_2=Q$ (see Fig.\ \ref{fig:htlexample}).
Our result for the retarded three-point function
agrees with the standard one;
we have also verified the four-point function.
Our result \Ref{ex:gammaaar} for the vertex function with
two $a$ indices agrees with that of \cite{fueki01},
Eq. (III.18e) (one must take into account that
their $\Gamma_{RRA}$ function is $2i$ times our $\Gamma_{aar}$,
that their momenta are incoming, and their
metric is different from ours.)

\section{Discussion}
\label{sec:discuss}

In this paper we have given the generating functional for
all hard thermal loops in the real-time formalism, \Ref{gena} and \Ref{genaa}.
The form of these amplitudes is deceptively simple:
a classical plasma physicist,
instructed of the fact that nonabelian charges
tend to precess in the presence of a gauge field, would have known
enough to simply write them down decades ago.
Indeed, we find that there are only two kinds of HTL amplitudes:
amplitudes with only one external Keldysh $a$ index, which describe
the (nonlinear) polarizability of a medium of nonabelian point-charges,
and amplitudes with two external Keldysh $a$ indices, which represent
current-current correlations in such a medium.
What we think is most interesting about our findings, is the fact
that these functions form the complete real-time HTL theory.
The essential reason for this, discussed in section \ref{sec:power},
is that the Keldysh $a$ (``difference'') fields,
compared to the Keldysh $r$ (``average'') fields, are unable to take
advantage of the large occupation numbers of the soft gauge
fields, hence diagrams containing more Keldysh $a$ indices
naturally tend to be subleading.
Incidentally, loop amplitudes with more than two external $a$ indices are
actually soft-dominated, not hard-dominated.

We have given simple effective Feynman rules, in section \ref{sec:rules},
which generate the hard thermal loop amplitudes; physically
these rules are nothing but a graphical representation of the nonabelian Vlasov
equations (see e.g. \cite{classical} or the review \cite{blaizot01}).
Diagrams not involving the cut propagator (f) of
Fig.\ \ref{fig:htlrules},
account for the classically induced current \Ref{j0ind2} due to a
background mean field, and whenever this (f) propagator appears, its role
is to account for the Gaussian fluctuations \Ref{fluctaa} of the
particle distribution functions (corresponding to fluctuations of
the ``$W$'' fields in the language of \cite{blaizot01}).
The importance of such Gaussian fluctuations was discussed previously
in the context of the hot electroweak theory (see \cite{huetson});
the analysis of the HTL theory given in \cite{Son},
of which we became aware after this work was completed, bears much
similarity with ours.

Our findings extend in a straightforward manner to nonequilibrium
setups: one should simply replace
the distribution functions in \Ref{systemarr1} and
\Ref{fluctaa} by their time-dependent expressions.
As was argued for in great detail in \cite{nonequilibrium},
this procedure will be correct provided the naive criterion
for it to make sense is satisfied: the distribution
functions of the particles should be slowly varying on the
length and time scale set by the ``soft'' gauge
fields (those for which the HTL effects become important).
Also, the perturbation theory employed here should make sense:
this requires a separation of scales between the ``soft'' scale and
the momenta of the particles
which dominate the integrals \Ref{systemarr1} and \Ref{fluctaa}
(in this paper this separation is $gT\ll T$).

Even though the close relationship between the HTL theory and classical
plasma physics has been widely recognized for a long time,
it was not clear, at least to the author, how this
understanding could be exploited in the context of
the next-to-leading order
calculation of dynamical quantities like the ones
listed in the introduction.
The difficulty is that at leading order one
has to deal not only with soft, classical physics,
but also with some truly quantum physics.
This is the context in which we believe
a systematic, \emph{a priori} fully quantum
mechanical approach, starting from the Feynman diagrams of the real-time
formalism, can be most useful. 

\section{Acknowledgements}
This work was supported in part by the Natural Sciences and Engineering
Research Council of Canada.

\end{document}